\documentclass[aps,prl,reprint,superscriptaddress]{revtex4-1}
\usepackage{amsmath,amssymb,bm}
\usepackage{graphicx}
\usepackage{hyperref}
\usepackage{braket}

\usepackage{xcolor}

\DeclareMathOperator{\Tr}{Tr}

\renewcommand\Im{\operatorname{Im}}

\begin{document}

\title{From Linear to Nonlinear Responses of Thermal Pure Quantum States}


\author{Hiroyuki Endo}
\author{Chisa Hotta}
\email{chisa@phys.c.u-tokyo.ac.jp}
\affiliation{Department of Basic Science, The University of Tokyo, 3-8-1 Komaba, Meguro, Tokyo 153-8902, Japan}
\author{Akira Shimizu}
\email{shmz@as.c.u-tokyo.ac.jp}
\affiliation{Department of Basic Science, The University of Tokyo, 3-8-1 Komaba, Meguro, Tokyo 153-8902, Japan}
\affiliation{Komaba Institute for Science, The University of Tokyo, 3-8-1 Komaba, Meguro, Tokyo 153-8902, Japan}

\date{\today}

\begin{abstract}
We propose a self-validating scheme to calculate the unbiased responses 
of quantum many-body systems to external fields of {\it arbibraty strength} at any temperature.
By switching on a specified field to a thermal pure quantum state of an {\it isolated} system, 
and tracking its time evolution, 
one can observe an intrinsic thermalization process driven solely by many-body effects.
The transient behavior {\it before thermalization} contains rich information on excited states, 
giving the linear and nonlinear response functions at all frequencies.
We uncover the necessary conditions to clarify the applicability of this formalism, 
supported by a proper definition of the {\it nonlinear} response function. 
The accuracy of the protocol is guaranteed by a rigorous upper bound of error exponentially 
decreasing with system size, and is well implemented in the simple ferromagnetic Heisenberg chain, 
whose response at high fields exhibits a nonlinear band deformation. 
We further extract the characteristic features of excitation of the
spin-$1/2$ kagome antiferromagnet; the 
wavenumber-insensitive linear responses from the possible spin liquid ground state, 
and the significantly broad nonlinear peaks 
which should be generated from numerous collisions of quasi-particles, 
that are beyond the perturbative description. 
\end{abstract}



\maketitle
When studying the dynamics of quantum many-body systems, one often encounters problems 
to which the linear response (LR) theory does not apply \cite{Kubo,KTH,FS2016,SF2017}. 
The LR in a weak external field gives us information only on the first-order excitations. 
For stronger fields, a nonlinear response (NLR) arises from higher-order processes, 
such as multi-photon and Raman processes, 
which provide us with abundant information on the excitations of the system \cite{Shen,Gibbs,Haug,Fujii,Ugeda}. 
Even a non-perturbative effect such as the band-gap renormalization is observed in the NLR \cite{Haug,Ugeda}.

Applications of the LR includes the determination of the fluctuations at or near equilibrium \cite{Kubo,KTH,FS2016,SF2017}, 
which is used to estimate the noises in electrical circuits \cite{Buckingham}. 
On the top of that, the NLR covers a wider
range of phenomena including harmonic generation \cite{Shen}, 
squeezing \cite{QObook1}, generation of entangled states \cite{Edamatsu}, 
and quantum computation \cite{QC1}. 
Therefore, in quest for appropriate nonlinear materials, 
basic understanding of the NLR is demanded. 

Unfortunately, it is hard to calculate or predict NLR in many-body quantum systems 
except for very limited cases, such as in an off-resonant field whose effects can be renormalized into the system parameters \cite{Inoue,Floquet}. 
This situation stands in contrast to the LR, 
with many elaborate methods developed, 
such as 
DMRG 
\cite{DDMRG1,DDMRG2,tDMRG}, 
quantum Monte Carlo simulations \cite{QMC1,QMC2}, 
kernel polynomial method \cite{KPM}, 
time correlation in pure quantum states 
\cite{pure1,pure2,Pastawski,Herbrych,pure3,pure4,pure5,Stainigeweg14,pure6,pure7},
and matrix-product state \cite{MPS}. 
Some of them were applied to NLR \cite{Pastawski,pure7} but only in 
a limited situation such as infinite temperature. 

If the system had only a few degrees of freedom, 
it would require a bath in order to `thermalize' after the quench. 
For such cases, numerical methods were successfully developed \cite{qj1,qj2,qj3}, 
where the LR and NLR would depend explicitly on the system-bath coupling. 
However, recent studies revealed that a many-body quantum system thermalizes even when isolated,  
provided that the number of conserved quantities is small 
\cite{vonN,Berry,Trotzky,Deutsch,Srednicki,Tasaki,Rigol,Alessio}. 
We then expect that a series of pure states that appear during the nonequilibrium relaxation process 
includes abundant information on the intrinsic dynamics of the system. 

In this {\em Letter}, we build a general and systematic protocol to extract the responses 
from the LR to the NLR regime based on the typicality approach
\cite{vonN,SugitaJ,Popescu,Goldstein,SugitaE,Reimann2007,SS2012,SS2013,HSS2014}. 
Our method is applicable to general many-body quantum systems and at any temperature. 
We prove rigorously that 
the time evolution of the expectation value of any low-order polynomial of 
local observables agrees with that obtained from the time evolution of the Gibbs state, 
within an error exponentially vanishing with increasing system size. 
We also clarify the necessary conditions to legitimate 
our NLR functions. 
These two guarantee the fully controlled observation beyond the LR. 
As illustrations, 
we analyze the NLR to a helical magnetic field for 
the kagome antiferromagnet 
\cite{white11,depenbrock12,iqbal13,waldmann,chisa}
as well as for the ferromagnetic Heisenberg chain. 

{\em Initial equilibrium state.---}
Consider a many-body quantum system with the Hamiltonian $\hat{H}$,
initially ($t \leq 0$) in a thermal equilibrium. 
Such an equilibrium state can be represented by various types of pure quantum states
\cite{vonN,SugitaJ,Popescu,Goldstein,SugitaE,Reimann2007,SS2012,SS2013,HSS2014}. 
Here, we choose the {\it unnormalized} canonical thermal pure quantum (cTPQ) state \cite{SS2013} 
as an initial equilibrium state given by 
\begin{equation}
\ket{\beta,N}
= \sum_{\nu} z_\nu \exp[- \beta \hat{H}/2 ] \ket{\nu},
\label{eq:ketbeta}
\end{equation}
at inverse temperature $\beta=1/T$ (where $k_{\rm B} = 1$) 
and system size $N$, 
with an arbitrary orthonormal basis of the Hilbert space 
$\{ \ket{\nu} \}_\nu$, 
and a random complex number $z_\nu = (x_\nu + i y_\nu)/\sqrt{2}$ \cite{HSS2014}. 
%
A single cTPQ state gives the free energy by 
$\beta F(T,N) = -\ln \braket{\beta,N | \beta,N}$, 
and accordingly all the thermodynamic properties within an error exponentially decreasing in $N$ \cite{SS2012,SS2013,HSS2014}.

{\em Response to external field.---}
Let us switch on an external vector field $\bm{h}$ at $t=0$, 
\begin{equation}
\bm{h}(\bm x,t) = h \bm{n}(\bm x) \theta(t),
\label{eq:hxt}
\end{equation}
where $h>0$, $\max_{\bm x} |\bm{n}(\bm x)|=1$, $\bm x$ is a spacial coordinate, 
and $\theta(t)$ is the step function.
Suppose that $\bm{h}$ couples to the system with the interaction 
$
\hat{H}_{\rm ext} 
= - \sum_x \bm{h}(\bm x,t) \cdot \bm{\hat{s}}(\bm x)
= - h \hat{B} \theta(t)
$,
where $\bm{\hat{s}}(\bm x)$ is a local operator of the system, 
and 
$
\hat{B} := \sum_{\bm x} \bm{n}(\bm x) \cdot \bm{\hat{s}}(\bm x)
$.

As a response to $\bm h$, 
we focus on a certain observable $\hat{A}$, 
which is an additive quantity or, more generally, a low-order polynomial 
(such as a two-point correlation) of local observables \cite{condA}.
Its deviation from the initial equilibrium value 
is given by 
\begin{equation}
\Delta A(t)
=
\braket{\hat{A}(t)}_{\beta,N}
-
\braket{\hat{A}}_{\beta,N},
\label{eq:DeltaA}
\end{equation}
where 
$\braket{\bullet}_{\beta,N} 
:=
\braket{\beta,N | \bullet | \beta,N}/\braket{\beta,N | \beta,N}$,
$
\hat{A}(t)
=
\hat{U}^\dagger(t) \hat{A} \hat{U}(t)
$,
and, taking $\hbar = 1$, 
$
\hat{U}(t)
=
\exp[ - i (\hat{H} - h \hat{B})t ]
$.

Here, similarly to what is rigorously proved for 
$\braket{\hat{A}}_{\beta,N}$ in the cTPQ state \cite{SS2013}, 
we show that $\braket{\hat{A}(t)}_{\beta,N}$ converges in probability 
to the nonequilibrium value calculated from the Gibbs state $\hat{\rho}_\beta$,
$\braket{\hat{A}(t)}_{\beta,N}^{\rm ens}
=
\Tr [ \hat{\rho}_\beta \hat{A}(t) ]$. 
Its deviation from the Gibbs ensemble 
after dropping off smaller-order terms is evaluated as, 
\begin{align}
\!\!\!
D&[\hat{A}(t)]^2 :=
\overline{
( 
\braket{\hat{A}(t)}_{\beta,N}
-{\langle}\hat{A}(t){\rangle}^{\rm ens}_{\beta,N}
)^2}
\nonumber\\
&\leq
{
\langle (\Delta \hat{A}(t))^2 \rangle^{\rm ens}_{2\beta,N}
+(\langle \hat{A}(t) \rangle^{\rm ens}_{2\beta,N} 
- \langle \hat{A}(t) \rangle^{\rm ens}_{\beta,N} )^2
\over 
\exp [2\beta \{ F(T/2, N)-F(T, N) \}]
},
\label{eq:DAt}
\end{align}
where $\overline{\bullet}$ denotes average over realizations of 
$\{ z_\nu \}$,
and 
$\Delta \hat{A}(t) :=\hat{A}(t)
- \langle \hat{A}(t) \rangle^{\rm ens}_{2\beta,N}$.
For every finite $\beta$, 
$
F(T/2, N)-F(T, N) = \Theta(N)
$ 
\cite{ordersymbol}
because the entropy $S = - \partial F/\partial T =\Theta(N)$.
Hence, the denominator of the rhs of Eq.(\ref{eq:DAt}) is 
$e^{\Theta(N)}$. 
Now, if we consider a typical case where 
$\hat{A}$ is an $m$-degree polynomial of bounded 
local observables \cite{unbounded},
the numerator is bounded to 
$\leq \Theta(N^{2m})$. 
We thus find  
$D[\hat{A}(t)]^2 \leq \Theta(N^{2m})/e^{\Theta(N)}$, 
which becomes exponentially small with increasing $N$. 
According to a Markov type inequality, 
this implies that 
$\braket{\hat{A}(t)}_{\beta,N}$ 
converges to ${\langle}\hat{A}(t){\rangle}^{\rm ens}_{\beta,N}$  with probability exponentially close to one, 
as in the equilibrium case \cite{SS2012,SS2013,HSS2014}.
Therefore, Eq.~(\ref{eq:DeltaA}) gives the correct response of the system of size $N$ 
with exponentially small error. 

{\em Linear and nonlinear susceptibility.---}
The LR and NLR need to be discussed separately.
When $h$ is small enough, the response extrapolates to that 
obtained from the LR theory \cite{Kubo,KTH,FS2016,SF2017}. 
In this LR regime, 
the linear susceptibility (or admittance) $\chi(\omega)$, 
which is the Fourier transform of the LR function \cite{Kubo,KTH,FS2016,SF2017}, 
does not depend on the profile of $\bm{h}$ along the time axis. 
Therefore, it is sufficient to consider the specific time dependent profile
Eq.~(\ref{eq:hxt}), to obtain the general form of $\chi(\omega)$ 
as a function of frequency $\omega$.  
Assuming that $\hat{A}$ is an additive observable, we obtain the following formula 
\begin{equation}
\chi(\omega)
=
{\Delta A(+\infty) \over Nh}
- i \omega
\int_0^\infty {\Delta' A(t) \over Nh} e^{i \omega t} dt,
\label{eq:chi}
\end{equation}
where 
$\Delta' A(t) 
:= 
\braket{\hat{A}(t)}_{\beta,N}
-
\braket{\hat{A}(+\infty)}_{\beta,N}$.
According to Kubo \cite{Kubo}, 
$\chi(\omega)$ is explicitly given by the retarded Green function at  equilibrium, 
which contains the information on the elementary excitations, whose nature could thus 
be examined by evaluating $\braket{\hat{A}(t)}_{\beta,N}$ for sufficiently small $h$. 
One can further specify the wavenumber $\bm{q}$ in $\bm{h}$, 
in order to obtain the $\bm{q}$-dependent susceptibility 
$\chi(\bm{q}, \omega)$. 
These points will be illustrated shortly. 

At larger $h$, the correspondence with the LR theory breaks down. 
Still, 
we use Eq.~(\ref{eq:chi}) as the definition of the {\em nonlinear susceptibility} 
$\chi(\bm{q}, \omega; h)$ with explicit $h$-dependence, 
because it is well-defined even in this NLR regime 
and is continuously connected to the linear one. 

Here, we do not follow the conventional perturbative definition 
in nonlinear optics \cite{Shen}. 
Our $\chi(\bm{q}, \omega; h)$ could treat nonperturbative effects 
such as the nonlinear band deformation, as we see shortly. 

{\em Necessary conditions.---}
In actual physical systems, 
Eq.(\ref{eq:chi}) gives correct predictions 
provided that $\hat{H}$ and $\hat{H}_{\rm ext}$ are the realistic Hamiltonians \cite{SK,SF2017}. 
However, in model calculations,
the Hamiltonian is often too idealized, 
as in the case of integrable Hamiltonians obtained by neglecting small but nontrivial interactions. 
Usually such idealization does not affect the quality of the {\em equilibrium} properties, whereas,
it often happens that they give wrong predictions about
{\em nonequilibrium} properties \cite{SK,SF2017,SM}. 

To reasonably predict nonequilibrium properties of a system, 
the following conditions are necessary:
(i) $[\hat{A}, \hat{H} - h \hat{B}] \neq 0$
because otherwise $\hat{A}$ would not respond to $\bm{h}$ at all.
(ii) $[\hat{A}, \hat{H}] \neq 0$ and $[\dot{\hat{B}}, \hat{H}] \neq 0$, 
since otherwise the state would depend on $\bm{h}$ in the distant past, 
as explicitly shown in the LR regime \cite{Kubo,KTH,FS2016,SF2017}. 
(iii) In cases where $\hat{H} - h \hat{B}$ has equilibrium states 
\cite{haseq},
the equilibrium susceptibility $\chi_{\rm eq}$ 
should agree with the $\omega \to 0$ limit
of Eq.~(\ref{eq:chi}) 
apart from a small difference of $o(1)$ due to 
equilibrium fluctuations. 
If not, the result would be inconsistent with equilibrium statistical mechanics. 
Notice that the temperature rises 
from that of the initial state due to $\bm{h}$, 
and $\chi_{\rm eq}$ should be measured at that temperature. 
In the LR theory, by contrast, the temperature remains the same within the order of $\Theta(h)$. 
Hence, condition (iii) is a generalization of 
that of the LR theory \cite{SM} to the NLR regime.
%
%
%
%
%
These conditions (i)-(iii) 
and Eqs.~(\ref{eq:ketbeta})-(\ref{eq:chi})
constitute our protocol. 

{\em Numerical method.---}
We employ the cTPQ state \cite{SS2013}, $\ket{\beta,N}$, as the initial 
equilibrium state, and adopt the Chebyshev polynomials expansion to 
obtain $\hat{U}(t)$ \cite{Tal-Ezer}. 
This part dominates the total numerical cost, although much less costly than the full diagonalization. 
Throughout the time evolution,  
the state 
keeps its purity, unlike the systems coupled to baths 
\cite{qj1,qj2,qj3}.

Our protocol is almost self-validating in the sense that the upper bound of the error 
$D[\hat{A}(t)]$ in the rhs of (\ref{eq:DAt}) 
is evaluated within the protocol;
the denominator is calculated in a similar manner as above, 
and $F(T,N)$ is obtained from $\| |\beta,N\rangle \|$. 
Notice that for small $N$ and low $T$, $D[\hat{A}(t)]$ can become rather large, 
in which case, we average over $\cal{M}$-independent choices of $\{ z_\nu \}$ 
to reduce $D[\hat{A}(t)]$ by a factor of $1/\sqrt{\cal{M}}$. 
[We take ${\cal M}=20$ and $3$ in Figs.~\ref{fig2} and \ref{fig3}, respectively.]

\begin{figure}[tb]
\centering
\includegraphics[width=0.46\textwidth]{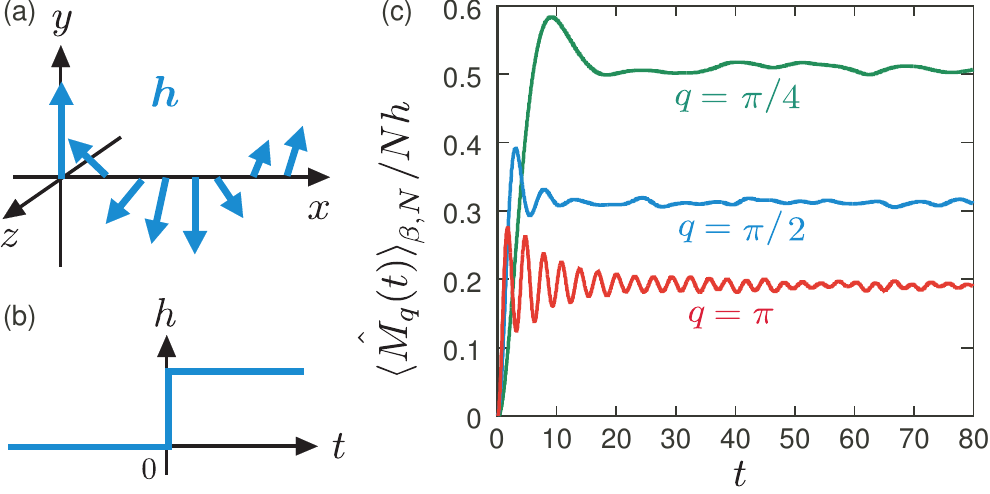}
\caption{
(a)(b) Schematic illustration of 
$\bm h(x,t)$. 
(c) Time evolution of 
$\langle \hat{M}_q (t) \rangle_{\beta,N}/Nh$, 
at $h=0.1$, $\beta=2$, and $N=16$.
}
\label{fig1}
\end{figure}

\begin{figure}[tb]
\centering
\includegraphics[width=0.49\textwidth]{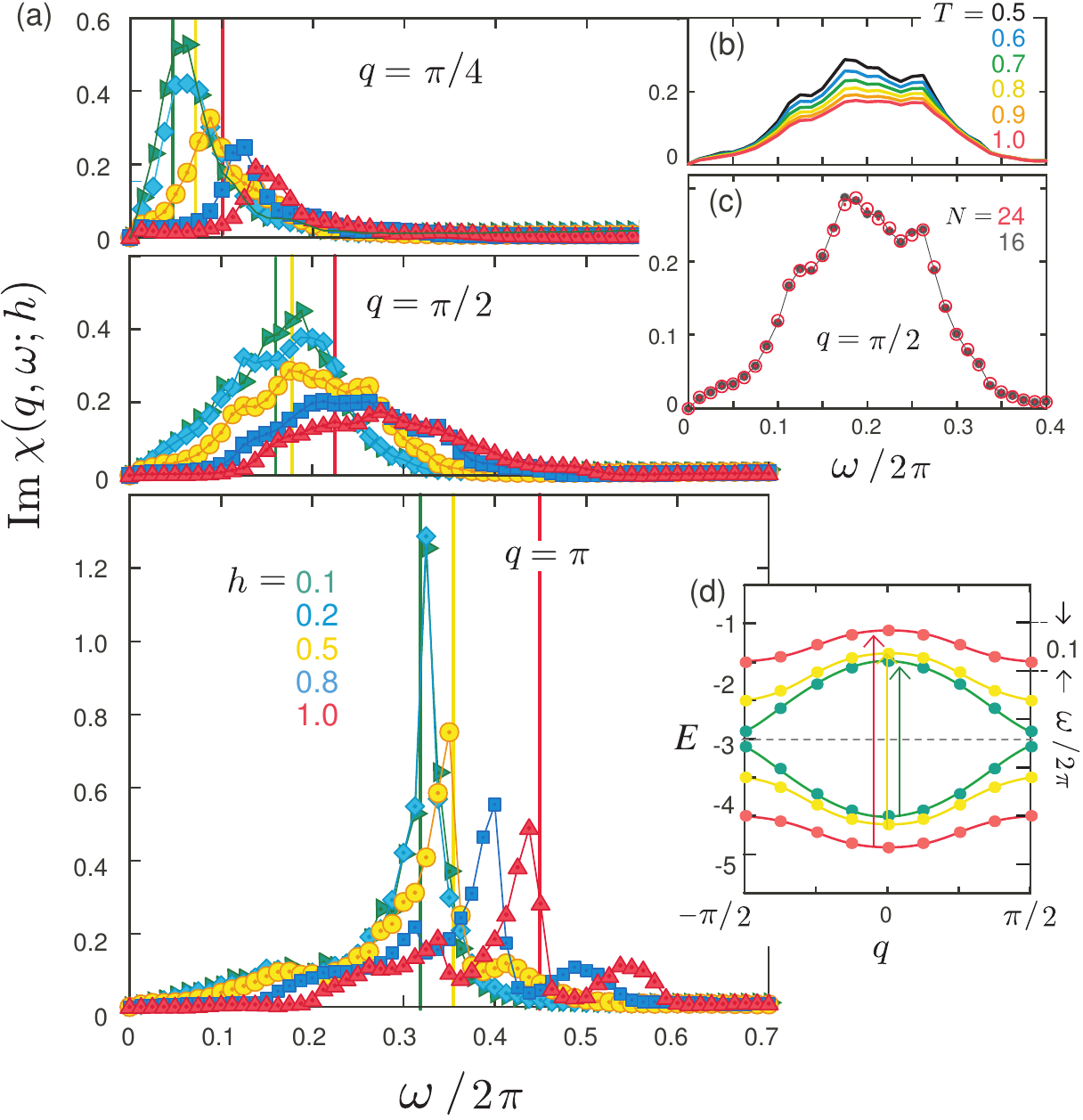}
\caption{
Results for the ferromagnetic Heisenberg chain. 
(a) $\Im\chi(q,\omega;h)$ at $T=0.5$
for $h=0.1$-$1.0$, 
$q=\pi/4$-$\pi$, $N=16$. 
(b) $\Im\chi(q,\omega;h)$ at $T=0.5$-$1.0$ for $h=0.5, q=\pi/2, N=16$.
(c) Comparison between $N=16$ and $24$ at $T=0.5$, $h=0.5$.
(d) Single-magnon dispersion 
in a magnetic field 
of $q=\pi$, 
whose transition energies 
indicated by the arrows 
are denoted by the vertical lines in (a) for $q=\pi$.} 
\label{fig2}
\end{figure}

\begin{figure}[tb]
\centering
\includegraphics[width=0.49\textwidth]{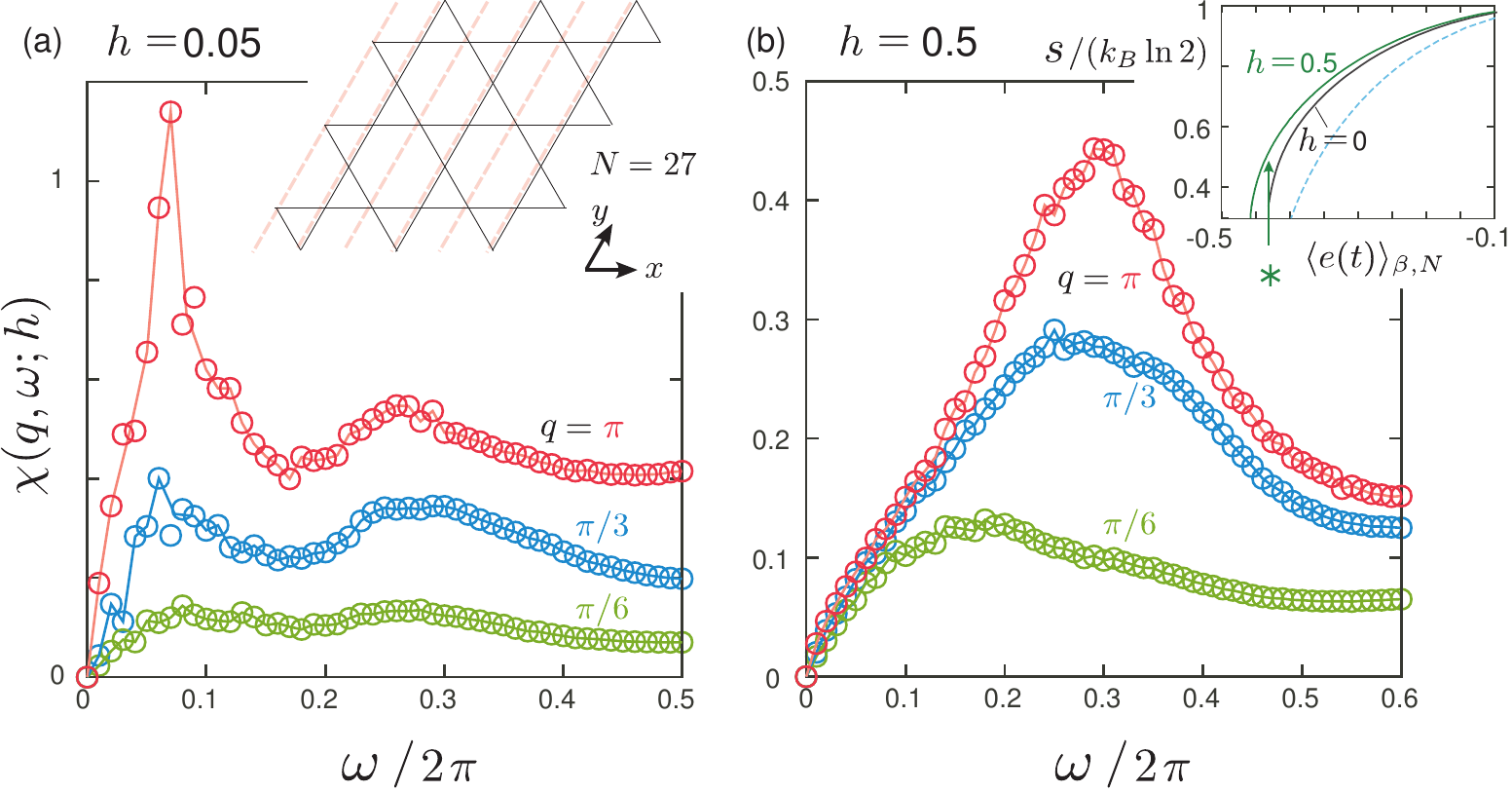}
\caption{
Results for the spin-1/2 kagome Heisenberg antiferromagnet. 
$\Im\chi(q,\omega;h)$ at $T=0.1$ and $N=27$ 
for (a) $h=0.05$, the LR regime 
and (b) $h=0.5$, the NLR regime. 
The inset shows the $N=27$ cluster. 
The inset of (b) is the entropy density $s/(k_B\ln 2)$ as a function of energy density $e$ 
obtained by the equilibrium microcanonical TPQ calculation \cite{SS2012}. 
Black and green solid lines are $h=0$ 
and $h=0.5$ ($q=\pi/3$) and the broken line 
is the case of ferromagnetic chain for comparison. 
The star is $\langle e(t)\rangle_{\beta,N}$ at the actual time evolution. 
}
\label{fig3}
\end{figure}

{\em Application to ferromagnetic Heisenberg chain.---}
We apply our protocol to the ferromagnetic Heisenberg chain, 
$\hat{H} = - \sum_{x} \hat{\bm s}(x) \cdot \hat{\bm s}(x+1)$, at 
$N=16$ and $24$ 
with the periodic boundary. 
Here, a uniform magnetic field would not satisfy 
the necessary condition (ii), 
Instead, we set $\bm{h} =h \bm{n}(x)$ as a helical magnetic field 
in the $y$-$z$ plane, i.e., 
$\bm{n}(x) = (0, \cos (qx), \sin (qx))$  with $q=n(2\pi/N)$ 
($n$: integer). 
The spatial and time-dependent profiles of $\bm{h}$ 
are shown in Figs.~\ref{fig1} (a) and (b), respectively.
Then we have
$
\hat{H}_{\rm ext} 
= - \sum_x \bm{h}(x,t) \cdot \bm{\hat{s}}(x)
= - h \hat{M}_q \theta(t)
$,
where $\hat{M}_q$ is the helical magnetization, 
\begin{equation}
\hat{M}_q :=
\sum_x
[\cos (qx) \hat{s}_y(x) 
+\sin (qx) \hat{s}_z(x)].
\end{equation}
We take $\hat{M}_q$ also as the observable of interest, $\hat{A}$, 
i.e., $\hat{A} = \hat{B} = \hat{M}_q$. 
Then, $\Delta A(t)  = \Delta M_q(t) =  \braket{\hat{M}_q(t)}_{\beta,N}$
since $\braket{\hat{M}_q}_{\beta,N}=0$ in the initial equilibrium state. 
The above setup satisfies 
all the necessary conditions (i)-(iii) 
((iii) has been confirmed numerically). 

Figure \ref{fig1}(c) shows the time evolutions of $\braket{\hat{M}_q(t)}_{\beta,N}/Nh$. 
It approaches a nearly constant value for every $q$, 
indicating the ``thermalization" \cite{thermalization1,thermalization2,thermalization3}. 
The transient behavior of time evolution {\it before thermalization} contains rich information on the low-energy excited states, 
which is reflected in $\chi(q, \omega; h)$. 

Here, we focus on its imaginary part, $\Im \chi$, 
whose peak in the LR regime indicates elementary excitations. 
To guarantee the convergence of $\Im \chi$, 
we take a long enough time-window, $t_{\rm end}=80$-$160$. 
We further calculate the round-trip evolution 
$\ket{\beta',N} := 
\hat{U}(-t_{\rm end}) \hat{U}(t_{\rm end}) \ket{\beta,N}$, 
which should equal $\ket{\beta,N}$ 
if the time evolution is correctly carried out. 
For a time step $\Delta t=1/50$ and the Chebyshev polynomials up to 500th order, 
the fidelity becomes 
$|\braket{\beta,N | \beta',N}|^2/\braket{\beta,N | \beta,\!N} \braket{\beta',\!N | \beta',\!N}
=1 \pm 3 \times 10^{-15}$ \cite{accuracy}. 
This confirms the perfect accuracy of our time evolution.

The highlight of the present protocol is the unbiased evaluation of both LR and NLR. 
The obtained $\Im \chi(q, \omega; h)$ 
for $\beta=2$ are plotted in Fig.~\ref{fig2}(a) for $q=\pi/4, \pi/2$ and $\pi$.
At $h\gtrsim 0.1$ the peaks of the spectra show significant shift 
and broadening, which is a strong nonlinear effect. 
At lower (higher) temperature, the peaks and dips of 
$\chi(q, \omega; h)$ become sharper (broader), 
as shown in Fig.~\ref{fig2}(b).
This happens because spins become more 
paramagnetic and thus less sensitive to $h$ at higher $T$. 
Since the finite-size effects are negligibly small (Fig.~\ref{fig2}(c)), 
we concentrate on the case of $N=16$. 

{\em LR regime of Heisenberg chain.---}
When $h \lesssim 0.1$, 
the response does not depend on $h$;  
$\chi(q, \omega; h) \to \chi(q, \omega)$ (see Supplemental Fig.~S1). 
In this LR regime, 
$\chi$ agrees with the Kubo formula (we confirmed for $N=8$)
except that the peaks and dips are broadened by 
$\delta \omega \sim 1/ t_{\rm end}$ 
because of the finite interval $0 \leq t \leq t_{\rm end}$ 
in the Fourier transformation. 

The fully polarized ground state of this model hosts a series of magnon excitations 
\cite{Dyson1,Dyson2,Wortis,Longo,Fukuhara}. 
One can construct a small subspace that is 
spanned by the zero-, one-, and two-magnon states \cite{SM}.
By directly applying the Kubo formula to this subspace,
we obtain 
$\Im \chi_{\rm sub}$ \cite{SM},
which is consistent with our $\Im \chi$;
the microscopic origin of the peaks at $h \le 0.1$ is identified as 
the transitions from a few lowest one-magnon levels to the higher ones, 
as well as to the continuum \cite{SM}. 
This kind of treatment works to clarify the physical origin of $\chi$, 
but is usually not available, e.g. in the kagome antiferromagnet we see shortly.

{\em NLR regime of Heisenberg chain.---}
The spectrum at $h \gtrsim 0.1$ 
in Fig.\ref{fig2}(a) shows a shift and the significant modification in its shape. 
Our protocol properly captures these nonlinear effects clearly beyond the scheme of the Kubo formula. 
Here, the term $- h \hat{M}_q$ cannot be treated as a small perturbation, 
and hence, to interpret the NLR, 
we diagonalize the full Hamiltonian $\hat{H} - h \hat{M}_q$ 
in the subspace we used to interpret the LR 
\cite{thesis}. 
The single magnon dispersion (solid line in the Supplemental Fig.~S1) 
is then folded by the period of $q$,  
and the gap opens at $k=\pi/q$, resulting in a band deformation as shown in Fig.~\ref{fig2}(d).
Then, the transition energy between the subbands at $k=0$, 
corresponding to the peak position, increases with $h$. 
Thus, this picture explains semi-quantitatively 
the nonlinear peak shift observed in Fig.~\ref{fig2}(a), 
validating Eq.~(\ref{eq:chi}). 
However, the complete NLR spectra,
beyond such a simple picture, 
is disclosed for the first time by our protocol.

{\em Kagome antiferromagnet. ---} 
We now present the dynamical responses of the spin-1/2 kagome antiferromagnet 
that had been unreachable in any of the previous techniques. 
The model is considered to host a spin liquid ground state 
\cite{white11,depenbrock12,iqbal13}, 
and a densely populated low-lying nonmagnetic excitations
\cite{waldmann}. 
Figures \ref{fig3}(a) and \ref{fig3}(b) show $\Im\chi(q,\omega;h)$ 
in the LR ($h=0.05)$ [see Supplementary Fig.S2(b)] 
and NLR ($h=0.5$) regimes, respectively.
Here, we apply a magnetic field $\bm h=h(0,\cos(qx),\sin(qx))$, 
varying along the $x$-direction with $q=n\pi/3$ 
while uniform in the $y$-direction on an $N=27$ cluster \cite{N=8}. 
Then the necessary conditions (i)-(iii) are satisfied.

The LR distinctly differs from Fig.~\ref{fig2} 
in that the three different $q$'s all show very similar profiles 
(except for the peak height), i.e., a characteristic wavenumber is absent. 
This seems to share a common context to the featureless magnetic 
structure factors of the frustrated spin liquid Mott insulator \cite{mizusaki06}. 
We also find that the first peak exists at around $0.05$, in consistency with the 
position of the spin gap \cite{chisa}, if present. 

In the NLR regime, a significantly broad peak is found. 
In the presence of strong many-body effects, 
the number of collisions among correlated particles generated by the strong field increases rapidly, dominating the NLR. 
In this case, the perturbative descriptions~\cite{Kubo,Shen} break down. 
We expect this to happen in the present model due to large 
entropy density $s$; 
in the time evolving pure state at $q=\pi/6$, 
it actually amounts to $s \sim (k_B\ln 2)/2$, 
half of the total value (inset of Fig.~\ref{fig3}(b)). 

{\em Concluding remarks.---}
If one simply replaces the observable, $\hat A$, of the TPQ formulation \cite{SS2012,SS2013,HSS2014}
with the Heisenberg operator, $\hat A(t)$, 
it easily yields wrong predictions on the LR,
unless some conditions are fulfilled \cite{KTH,SK,SF2017}. 
Undoubtedly, this problem becomes more serious for the NLR.
We provided a solution to this fundamental problem by 
identifying the {\it necessary conditions (i)-(iii)}. 
It works hand in hand with 
the proper definition of the nonlinear susceptibility Eq.(\ref{eq:chi}) 
which has a nonperturbative form, 
and the necessary conditions serve as a safeguard to avoid unphysical results. 
In the limit of weak fields, our susceptibility and the necessary conditions 
recover those of the LR theory. 

On the numerical side, 
our protocol itself has neither restrictions on the system size 
(except for a limitation by available numerical resources) 
nor the types of models, 
regardless of how rapidly the entanglement grows in time evolution. 
So far, there had been no guarantees in both the LR and NLR for larger-scale approximate calculations. 
Our protocol provides a reliability check within an available system size beforehand. 
By computing the response function for the kagome antiferromagnet, we proved 
that our method is well founded even in one of the most 
challenging models in condensed matter. 
\begin{acknowledgments}
We thank R. Hatakeyama, R. Hamazaki and K. Asano
for helpful discussions,
and J. Romhandyi for critical reading of the manuscript.
This work is supported by 
JSPS KAKENHI Grant Numbers 
JP15H05700, JP26287085, 
JP17K05533, JP18H01173, JP17K05497, and JP17H02916.
\end{acknowledgments}

\clearpage

\pagestyle{empty}
\parindent=0mm
\setlength{\textwidth}{1.15\textwidth}

\begin{figure*}
\vspace{-10mm}
\hspace{-28mm}
\includegraphics[page=1,clip]{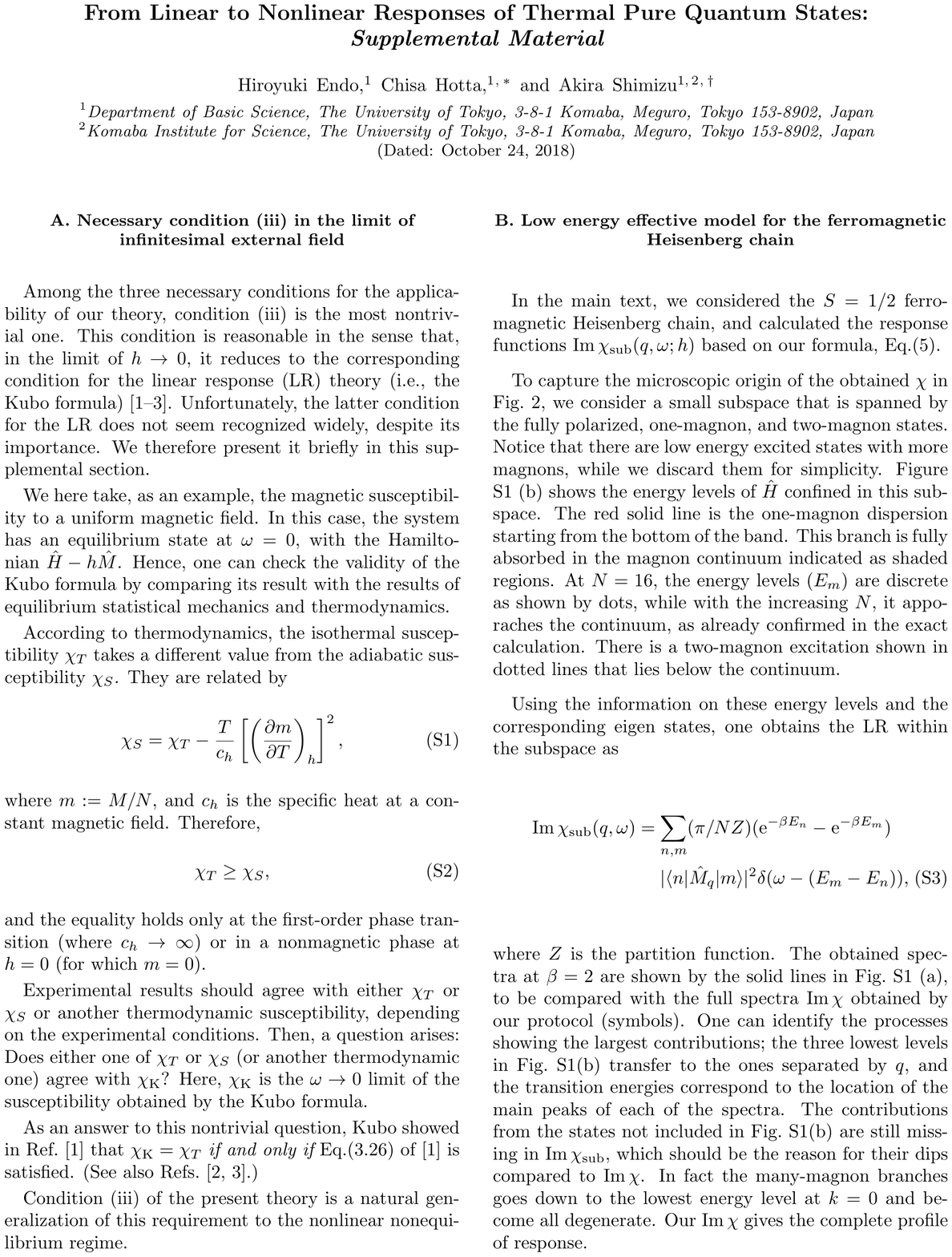}
\end{figure*}

\begin{figure*}
\vspace{-10mm}\hspace{-28mm}
\includegraphics[page=2,clip]{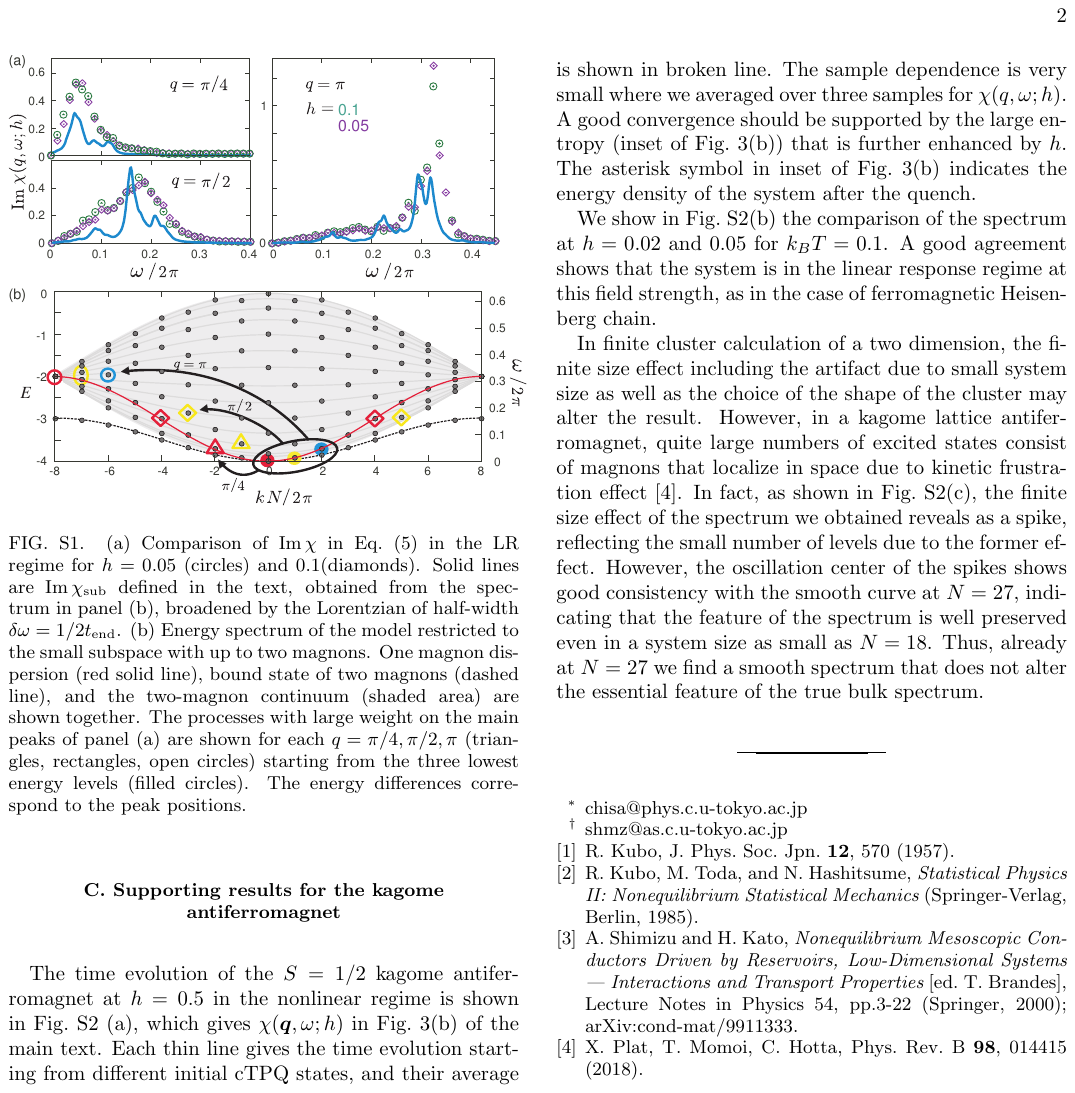}
\end{figure*}

\begin{figure*}
\vspace{-10mm}\hspace{-28mm}
\includegraphics[page=3,clip]{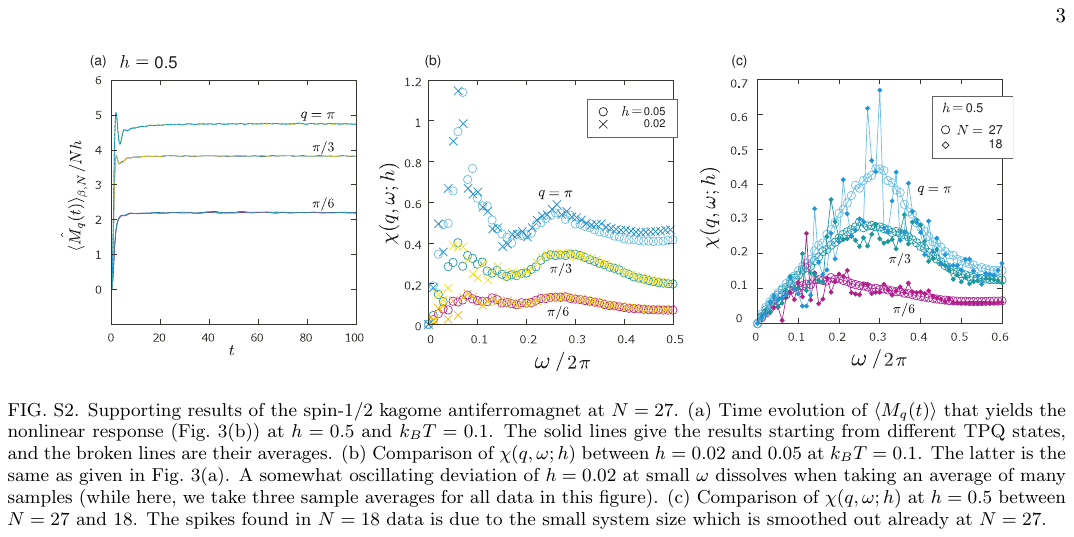}
\end{figure*}


\begin{thebibliography}{99}
\bibitem{condA}
This condition on $\hat{A}$ can be relaxed to any observables 
whose norm is bounded by a polynomial of $N$.
\bibitem{ordersymbol}
%
For a function $g(N)$, we say 
$g = \Theta(N^k)$
if $g/N^k$ approaches a positive constant as $N \to \infty$.
\bibitem{unbounded}
Physically, the same result is expected for the case of unbounded local observables.
\bibitem{haseq}
Although $\bm{h}$ is static for $t \geq 0$, 
it does not necessarily mean that 
$\hat{H} - h \hat{B}$ has equilibrium states, 
e.g., $h \bm{n}(x)$ may be a uniform electric field
(and $\bm{s}(x)$ is the local charge),
in which equilibrium states are impossible.
\bibitem{accuracy}
In practice, 
looser condition can still give accurate results depending on the choice of the system. 
For the kagome antiferromagnet, we take $\Delta t=0.2$, $t_{\rm end}=100$, and the Chebyshev polynomials up to 100th order, 
which are enough to 
keep the fidelity loss as small as $\sim 10^{-10}$.
\bibitem{N=8}
We confirmed that an $N=18$ cluster also shows a very similar 
broad-peak structure,
with the same peak position, so that the finite size effect does not alter
the result.
\end{thebibliography}
\end{document}